\documentclass[aps,prc,twocolumn,showpacs,byrevtex,amsmath,amssymb,nofootinbib]{revtex4}

\usepackage{graphicx}% Include figure files
\usepackage{dcolumn}% Align table columns on decimal point
\usepackage{bm}% bold math
\usepackage{hyperref}% add hypertext capabilities
\usepackage{comment}
\usepackage[normalem]{ulem}

\def\nuc#1#2{\relax\ifmmode{}^{#1}{\protect\text{#2}}\else${}^{#1}$#2\fi}
\newcommand{\etal}{\textit{et al.~}}

\newcommand{\vecr}{{\vec r}}

\newcommand{\be}{\begin{eqnarray}}
\newcommand{\ee}{\end{eqnarray}}
\newcommand{\dif}{{\mathrm{d}}}
\newcommand{\bwt}{\begin{widetext}}
\newcommand{\ewt}{\end{widetext}}

\newcommand{\threejz}[3]{\begin{pmatrix}#1&#2&#3\\0&0&0\end{pmatrix}}
\newcommand{\sixj}[6]{\begin{Bmatrix}#1&#2&#3\\#4&#5&#6\end{Bmatrix}}
\newcommand{\ninej}[9]{\begin{Bmatrix}#1&#2&#3\\#4&#5&#6\\#7&#8&#9\end{Bmatrix}}
\newcommand{\reduced}[3]{\langle#1 \| #2 \| #3\rangle}
\newcommand{\coords}{\vec{R},\vec{r},\xi}

\begin{document}

\title{Continuum-Discretized Coupled-Channels calculations with core excitation} 


%\author{}

\author{R.~de Diego}
\email{raulez@cii.fc.ul.pt}
%\altaffiliation{Present address: Centro de F\'{\i}sica Nuclear,
%Universidade de Lisboa, Av. Prof. Gama Pinto 2, P-1649-003 Lisboa,
%Portugal}% 
\affiliation{Departamento de FAMN, Facultad de F\'{\i}sica, Universidad de Sevilla, Apdo.~1065, E-41080 Sevilla, Spain\\
Centro de F\'{\i}sica Nuclear, Universidade de Lisboa, Av. Prof. Gama Pinto 2, P-1649-003 Lisboa, Portugal}%

\author{J.~M.\ Arias}
\email{ariasc@us.es}%
\affiliation{Departamento de FAMN, Facultad de F\'{\i}sica, Universidad de Sevilla, Apdo.~1065, E-41080 Sevilla, Spain}

\author{J.~A. Lay}
\email{lay@us.es}
\altaffiliation{Present address: Dipartimento di Fisica e Astronomia, Universit\`{a} di Padova, I-35131 Padova, Italy and 
 Istituto Nazionale di Fisica Nucleare, Sezione di Padova, I-35131 Padova, Italy}%
\affiliation{Departamento de FAMN, Facultad de F\'{\i}sica, Universidad de Sevilla, Apdo.~1065, E-41080 Sevilla, Spain}

\author{A.~M.\ Moro}
\email{moro@us.es}
\affiliation{Departamento de FAMN, Facultad de F\'{\i}sica, Universidad de Sevilla, Apdo.~1065, E-41080 Sevilla, Spain}

\vspace{1cm}

\date{\today}

%\thanks{}
%\affiliation{}

\begin{abstract} 
The effect of core excitation in the elastic scattering and breakup of
a two-body halo nucleus on a stable target nucleus is studied. The
structure of the weakly-bound projectile is described in the
weak-coupling limit, assuming a particle-rotor model. The
eigenfunctions and the associated eigenvalues are obtained by
diagonalizing this Hamiltonian in a square-integrable basis
(pseudo-states). For the radial coordinate between the particle and
the core, a transformed harmonic oscillator (THO) basis is used. For
the reaction dynamics, an extension of the Continuum-Discretized
Coupled-Channels (CDCC) method, which takes into account dynamic core
excitation and deexcitation due to the presence of non-central parts
in the core-target interaction, is adapted to be used along with a pseudo-states (PS) basis.  
\end{abstract}

%\pacs{24.10.-i, 24.10.Eq, 25.10.+s, 25.45.De, 25.60.Gc}
 % summers
%\keywords{}

\maketitle

%-------------------------------------------------------------------
\section{\label{intro} Introduction}
%------------------------------------------------------------------

%importance of the continuum-------------
Nuclei in the proximity of the proton and neutron drip-lines are often
weakly bound, or even unbound, and hence their  properties are
influenced by positive-energy states. Collisions of these systems with
stable nuclei will also be influenced by the coupling to the unbound
states.  This effect was first noticed 
in deuteron-induced reactions, and later observed in the scattering of
other loosely bound nuclei, such as halo nuclei. Several formalisms
have been developed to account for the effects of the coupling to
breakup channels on reaction observables:  
Continuum-Discretized Coupled-Channels (CDCC) method
\cite{Raw74,Aus87}, the adiabatic approximation \cite{Ban00,Tos98},
the Faddeev/AGS equations  \cite{faddeev60,Alt},
and a variety of  semi-classical approximations
\cite{Typ94,Esb96,Kid94,Typ01,Cap04,Gar06}.   

%importance of core excitation and mixing--------------
Typically, these approaches make use of a few-body description of the
weakly bound nucleus.  Furthermore, in their standard formulations,
the constituent fragments are considered to be inert and, therefore, possible
excitations of them are ignored. This is a good  approximation for
deuteron scattering, for which both constituents can be considered
inert at the  energies of interest in nuclear studies, but it is
questionable for more complex systems. Moreover, bound and unbound
states of the few-body system are considered to be well described by
pure single-particle configurations. This approximation ignores
possible admixtures of different core states in the wave functions of
the complete projectile. These admixtures are known to be important,
particularly in the case of well-deformed cores, as for example in the
$^{11}$Be halo nucleus. 

%previous works and limitations---------------------------------
%dwba
In this work, we concentrate in two-body weakly bound nuclei composed
by a core plus a valence particle. For such systems, core excitation
effects in elastic breakup have been recently studied with an
extension of the Distorted Wave Born Approximation (DWBA) formalism which includes them within a
no-recoil approximation \cite{Cre11,Mor12,Mor12b}, referred to
hereafter as no-recoil XDWBA. These  calculations have shown that core
excitation effects  have a sizable influence in the magnitude of the
breakup cross sections \cite{Cre11,Mor12}. Moreover, these core
excitation effects interfere with the valence excitation mechanism,
altering the diffraction pattern in the resonant breakup angular
distributions \cite{Mor12b}. This method, being based in the Born
approximation, ignores higher order effects (such as
continuum-continuum couplings) and cannot be applied to describe the
effect of breakup on elastic scattering.  

%adiabatic
The effect of core excitation in elastic scattering has also been
studied \cite{Hor10}, using an extension of the adiabatic model of
Ref.~\cite{Ron97}. The formalism was applied to $^8$B+$^{12}$C, and
some contributions due to $^7$Be core excitations were found at large
angles. Due to the use of the adiabatic approximation this method is,
however, restricted to intermediate and high energies.  

%xcdcc
 A recent attempt to incorporate core excitation effects within a
 full-fledged coupled-channels calculation was done in
 Ref.~\cite{Summers06}, using an extended version of the CDCC formalism 
(XCDCC). %, and applied to the scattering of
% one-neutron halo nuclei. To account for the core-excitation
% mechanism, the valence-core and core-target interactions were
% described in terms of deformed potentials, assuming a rotor model for
% the core.
 The method was applied to
 the scattering of one-neutron halo nuclei, using deformed
 valence-core and core-target potentials to account for the
 core excitation mechanism.
 These calculations suggested a very small effect of
 core excitation,\footnote{Some equation errors and bugs in the
   programming of Ref.~\cite{Summers06} (and subsequent related
   papers) were detected during the control checks of the present work. An
   erratum to Ref.~\cite{Summers06} has already been produced by its
   authors~\cite{Sum14,Sum14b}. See also comment in Sect. IV.A.} in
 contrast with the results of Refs.~\cite{Cre11,Mor12,Mor12b}. 
In this work, we revisit the formulation of the XCDCC method of
Ref.~\cite{Summers06} and perform calculations for the elastic
scattering and breakup of $^{11}$Be on several targets at low and
intermediate energies. The aim of this work is to provide an improved
description of the reaction dynamics, as compared to the no-recoil
DWBA method and also to pin down the effect of core excitation in
elastic scattering and breakup. Our description of the reaction dynamics follows
closely the derivation of Ref.~\cite{Summers06}, but a new code to
compute the required coupling potentials has been developed in order
to provide an independent assessment of the importance of core
excitation effects in the scattering of halo nuclei.  
% Also, we aim at sheding some light toward the understanding of the
% importance of core excitation effects in the scattering of
% weakly-bound nuclei. 
The main difference between our approach and that of
Ref.~\cite{Summers06} relies on the description of the states of the weakly-bound projectile. 
 In \cite{Summers06}, the wave functions for these states were
obtained by direct integration of the multi-channel Schr\"odinger
equation, subject to the appropriate boundary conditions for bound or
unbound states. The latter were then grouped into bins,  
%these states were described within the particle-rotor model
%\cite{BM}. The  the unbound states of the compound  system were
%described 
% by continuum bins which, following the standard procedure,  
constructed by superposition of scattering  states, following the
standard average procedure.  
%These scattering states are obtained by direct integration of the
% Schr\"odinger equation, subject to the boundary conditions that
% incident waves occur for a given channel configuration, and outgoing
% waves appear for all channels compatible with the total angular
% momentum of the projectile.  
In this work, we use instead the so-called {\it pseudo-state} method, in which
the projectile states are approximated by the  
%For the description of the states of the two-body projectile, assumed
%to be composed by a valence coupled to a deformed core, we use the  
%THO method Ref.~\cite{Lay12}. In this method, the bound and unbound states of the system can be
eigenstates of the Hamiltonian in a truncated basis of
square-integrable functions. Negative-energy eigenvalues correspond to
the bound states of the system, whereas those located at positive  
energies, usually referred to as pseudo-states (PS), can be regarded
as a finite and discrete representation of the continuum spectrum. The
method has been successfully applied to two-  \cite{Mat03,Per02,Mor09}
and three-body problems \cite{Mat04a,Mat04b,Mat06,manoli08}.  
In particular, we make use of a Transformed Harmonic Oscillator
(THO) basis.  This basis has been  applied to
the case of spherical systems \cite{Mor09} and also to deformed
systems \cite{Lay12}. In both cases, the THO basis 
is used to describe the relative motion between the clusters and it is obtained by  
applying a Local Scale Transformation (LST) to the Harmonic Oscillator
(HO) basis. The LST, adopted from a previous  
work of Karataglidis \etal \cite{Amos}, is such that it transforms the
Gaussian asymptotic behavior into an exponential form,  thus ensuring
the correct asymptotic behavior for the bound wave functions.
The combined XCDCC+THO formalism is applied to %the reactions
$^{11}$Be+p, $^{11}$Be+$^{64}$Zn and $^{11}$Be+$^{208}$Pb reactions
and the effect of core excitation is discussed in each case (light,
medium and heavy target).

The work is structured as follows. In Sec.~\ref{sec:tho} we briefly
recall the THO basis used for the description of two-body systems with
core excitation. In Sec.~\ref{sec:xcdcc}, the XCDCC formalism,
particularized to our basis functions, is revisited. In
Sec.~\ref{sec:calc} the XCDCC+THO method is applied to several
reactions induced by the $^{11}$Be nucleus. Finally, in
Sec.~\ref{sec:summary} the main results of this work are summarized.

%--------schematic figure ---------------
\begin{figure}
{\par\centering \resizebox*{0.45\textwidth}{!}{\includegraphics{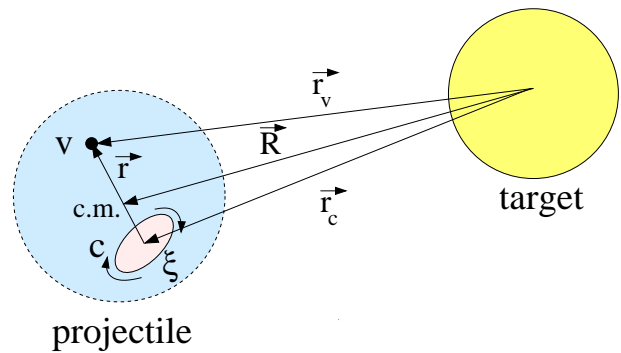}}\par}
 \caption{\label{fig:1} (Color online) Schematic sketch of the
   weakly-bound projectile composed by a core (c) and a valence
   particle (v). To study the scattering of the composite projectile with a inert target, within 
a three-body model, the relevant coordinates are the relative coordinate of the valence particle 
with respect to the core ($\vec r$) and that between the center of mass of the projectile and the target ($\vec R$). 
Note that the valence-target and core-target coordinates ($\vec r_v$ and $\vec r_c$, respectively) can be written in terms of $\vec r$ and $\vec R$
%The relevant degrees of freedom are the location of
%   the particle relative to the core ($\vec r$) and the internal
%   degrees of freedom of the core ($\xi$). 
%In order to study the   interaction of such a projectile with an inert target, the
%   positions of projectile and core centers of mass relative to the
%   target ($\vec R$ and $\vec r_c$, respectively) are needed.
}   
\end{figure}
%------------------------------------

%-----------------------------------------------------------------------
\section{\label{sec:tho}  Structure of the projectile in a  THO basis }  
%------------------------------------------------------------------------
In this section, we briefly review the features of the PS basis used
in this work to describe the states of a two-body composite
projectile, made of a valence particle (v) and a core nucleus (c) (see
schematic Fig.~\ref{fig:1}). The Hamiltonian of this system, $H_p$, is
described in the weak-coupling limit and is written as 
\begin{equation}
H_p(\vec{r},\xi)=T(\vec{r})+V_{vc}(\vec{r},\xi)+h_c(\xi),
\label{hproj}
\end{equation}
where $T(\vec{r})$ is the core-valence kinetic energy operator,
$V_{vc}$ is the valence-core interaction, and $h_c(\xi)$ is the
intrinsic Hamiltonian of the core. 
% The states of the system are computed by using an extension of the
% THO method proposed in Ref.~\cite{Lay12} to describe two-body
% composite systems but allowing core-excited admixtures in the
% description of their states and hence the possibility of dynamic
% core excitation mechanisms.  

In the calculations presented in this work, the composite system
(projectile) is treated within  the particle-rotor model
\cite{BM}. Therefore,  we assume that the core nucleus has a permanent
deformation  which, for simplicity, is taken 
to be axially symmetric. Thus, we can characterize the deformation by
a single parameter, $\beta_2$.  In the body-fixed frame, the  
surface radius is parameterized as $R(\hat{\xi})=R_0 [1 + \beta_2
  \, Y_{20}(\hat{\xi})]$, with $R_0$ an average radius. Starting from
a central potential, $V^{(0)}_{vc}(r)$, the full valence-core
interaction is  obtained by deforming this  interaction as,  
%In this model, the internal coordinates $\vec{\xi}$ describe the
%surface of the nucleus in the body-fixed frame. . 
\be
V_{vc}(\vec{r},\hat{\xi})=V_{vc}^{(0)}\left(r-\delta_2
Y_{20}(\hat{\xi})\right) , 
\label{Vvc}
\ee
with $\delta_2 = \beta_2 R_0 $, usually called deformation
length. Transforming to the space-fixed reference frame, and
expanding in  spherical harmonics, this {\it deformed} potential reads
(see e.g.~Ref.~\cite{Tam65}) 
%that can be casted in the form of Eq.~(\ref{vct2}) \cite{Tam65}
\be
V_{vc}(r,\theta,\phi)= \sqrt{4 \pi} \sum_{\lambda \mu} {\cal
  V}^{\lambda}_{vc}(r) {\cal D}^{\lambda}_{\mu 0}(\alpha,\beta,\gamma)
Y_{\lambda \mu}(\hat{r}) 
\label{vdef_vc}
\ee
with the radial form factors
\be
{\cal V}^{\lambda}_{vc}(r) =  \frac{\hat{\lambda}}{2} \int_{-1}^{1}
V_{vc}\left(r-\delta_2 Y_{20}(\theta',0)\right)  P_{\lambda}(u)~ \dif u , 
\label{vlam}
\ee
(with $u=\cos \theta'$ and $\hat{\lambda}\equiv
\sqrt{2\lambda+1}$). ${\cal D}^{\lambda}_{\mu 0}(\alpha,\beta,\gamma)$
is a rotation matrix, depending on the Euler angles
$\{\alpha,\beta,\gamma\}$ which define the transformation from the
body-fixed frame to the laboratory frame.  

% \be
% V_{vc}(\vec{r},\vec{\xi}) = \sqrt{4 \pi}  \sum_{\lambda \mu}V_{vc}^{\lambda}(r)  
% Y_{ \lambda \mu}(\hat{r}) {\cal D}^{\lambda}_{\mu 0}(\alpha,\beta,\gamma) 
% Y^{*}_{ {\lambda \mu} }(\hat{\xi})~,
% \label{vdef_vc}
% \ee 
% where the radial  form factors $V_{vc}^{\lambda}(r)$ are obtained by
% projecting the deformed potential (\ref{Vvc}) onto the  
% required multipoles. 

The eigenstates of the  Hamiltonian and their associated wavefunctions
can be obtained by solving a system of differential equations, as done
in Ref.~\cite{Summers06}. Alternatively, they can be obtained
diagonalizing the Hamiltonian matrix in a finite basis of square
integrable functions. In this work we use this second procedure. For
that, we choose a basis of the form: 
\begin{equation}
\phi^{THO}_{i,\alpha,J_p,M_p}(\vec{r},\xi)=R^{THO}_{i,\ell}(r) 
  \left[{\cal Y}_{(\ell s) j}(\hat{r}) \otimes \varphi_{I}(\xi) \right]_{J_p
  M_p} ,
\label{fTHO}
\end{equation}
where the label $\alpha$ denotes the set of quantum numbers $\{\ell,
s, j, I\}$, with $\vec{\ell}$ (valence-core orbital angular momentum) and
$\vec{s}$ (spin of the valence) both coupled to $\vec{j}$ (total valence particle
angular momentum). The total spin of the projectile, $\vec{J}_p$, is given by the
coupling between $\vec{j}$ and $\vec{I}$ (intrinsic spin of the core). The  
valence-core relative motion is described by the
functions $R^{THO}_{i,\ell}(r)$ (radial part) and ${\cal Y}_{(\ell s)
  j}(\hat{r})$ (spin-angular part), while the functions
$\varphi_{I}(\xi)$ describe the core states. The functions
$R^{THO}_{i,\ell}(r)$ are generated by applying a local scale
transformation (LST) to the spherical HO basis functions, 
%--------------------------------------------------------
\begin{equation}
\label{eq:tho}
R ^{THO}_{i, \ell}(r)= \sqrt{\frac{ds}{dr}}~ R^{HO} _{i, \ell}[s(r)],
\end{equation}
%--------------------------------------------------------
where $R ^{HO}_{i, \ell}[s(r)]$ (with $i=1,2,\ldots$) is the radial
part of the HO functions and $s(r)$ defines the LST. For the latter we
use the analytical prescription by Karataglidis \etal \cite{Amos}  
%---------------------------------------------------------------------
\begin{equation}
\label{lst}
s(r)  = \frac{1}{\sqrt{2} b} \left[  
 {\left(  \frac{1}{r} \right)^m  +  \left( \frac{1}{\gamma\sqrt{r}}
   \right)^m } \right]^{-\frac{1}{m}}\ , 
\end{equation}
%----------------------------------------------------------------------
that depends on the parameters $m$, $\gamma$ and the oscillator length
$b$. This transformation was shown in Ref. \cite{Amos} to depend weakly
on $m$. The value $m=4$ was proposed in \cite{Amos} and adopted
here. Thus, the adopted LST  depends on $\gamma$ and $b$. The ratio
$\gamma/b$ determines the range of the basis 
functions and the density of eigenstates as a function of the
excitation energy. As $\gamma/b$ decreases, the basis functions explore
larger distances and the corresponding eigenvalues concentrate at
smaller excitation energies.

The eigenstates of the Hamiltonian (\ref{hproj}) are expressed as
an expansion in the THO basis,  
\be
\Phi^{(N)}_{n,J_p,M_p}(\vec{r},\xi) = \sum_{i=1}^{N} \sum_\alpha
C^{n}_{i,\alpha,J_p} \phi^{THO}_{i,\alpha,J_p,M_p}(\vec{r},\xi), 
\ee
where $N$ is the number of radial functions retained in the truncated
THO basis, $n$ is an index identifying each eigenstate, and
$C^{n}_{i,\alpha,J_p}$ are the expansion coefficients of the
pseudo-states in the truncated basis. The sum in $i$ can be  performed to get
\be
\Phi^{(N)}_{n,J_p,M_p}(\vecr,\xi) = 
    \sum_{\alpha} \frac{u_{n,\alpha}^{J_p}(r)}{r} 
  \left[{\cal Y}_{(\ell s) j}(\hat{r}) \otimes \varphi_{I}(\xi) \right]_{J_p M_p}
%    \Phi_{\alpha,J_p,M_p}(\hat{r}, \xi) , 
\ee
with 
\be
u_{n,\alpha}^{J_p}(r) =  r \sum_{i=1}^{N} C^n_{i,\alpha,J_p}~  R^{THO}_{i,\ell}(r).
\ee
The negative eigenvalues of the Hamiltonian (\ref{hproj}) are
identified with the energies of bound states whereas the positive
ones correspond to a discrete representation of the continuum
spectrum.

%-----------------------------------------------------
\section{\label{sec:xcdcc} Scattering framework}
%-----------------------------------------------------
Once the projectile wave functions have been obtained, we proceed to
solve the three-body scattering problem. The formalism has been derived and
presented in detail in Ref.~\cite{Summers06} so we summarize here the
main formulae and adapt them to our PS scheme.  We express the three-body wave functions $\Psi_{J_T,
  M_T}$ in terms of the set $\{\Phi^{(N)}_{n,J_p}\}$: 

\begin{equation}
\Psi_{J_T, M_T}(\vec{R},\vec{r},\xi)=\sum_{\beta}
\chi_\beta^{J_T}(R)\left[Y_L({\hat{R}})\otimes\Phi^{(N)}_{n,J_p}(\vec{r},\xi)\right]_{J_T,M_T},   
\label{f3b}
\end{equation}
where, in addition to the projectile coordinates $\vec{r}$ and $\xi$,
we have the relative coordinate $\vec{R}$ between the projectile
center of mass and the target (assumed to be structureless), see
Fig. \ref{fig:1}. 
The different quantum numbers are labeled by $\beta=\{L, J_p, n\}$,
where $\vec{L}$ (projectile-target orbital angular momentum) and $\vec{J}_p$ both couple
to the total spin of the three-body system $\vec{J}_T$. The spin of the
target is ignored for simplicity of the notation.

The radial coefficients,  $\chi_\beta^{J_T}(R)$, from which the
scattering observables are extracted, are calculated  
by inserting (\ref{f3b}) in the Schr\"odinger equation, giving rise to
a system of coupled differential equations. The main physical
ingredients of these coupled equations are the coupling potentials: 
\begin{equation}
U_{\beta,\beta^\prime}^{J_T}(R)=\langle \beta;
J_T|V_{ct}(\vec{R},\vec{r},\xi)+V_{vt}(\vec{R},\vec{r})|\beta^\prime;
J_T\rangle , 
\label{cpot}
\end{equation}
where we follow the notation used in Ref.~\cite{Summers06},
\begin{equation}
\langle \hat{R}, \vec{r}, \xi |\beta; J_T\rangle =
\left[Y_L({\hat{R}})\otimes\Phi^{(N)}_{n,J_p}(\vec{r},\xi)\right]_{J_T}. 
\label{coupledbasis}
\end{equation}

%-------------------------------
The valence particle-target interaction ($V_{vt}$) is assumed to be
central, and will be represented by a phenomenological optical
potential describing the valence particle-target elastic scattering at
the appropriate energy per nucleon. On the other hand, the core-target
interaction is assumed to contain a non-central part, responsible for
the dynamic core excitation/deexcitation mechanism. In general, this
interaction can be expressed in the multipolar form: 
%Consistently with {\sc fresco} convention, the multipole expansion of
%the core-target interaction is assumed to have the form (see also
%\cite{Tho09}): 
\be
\label{vct}
V_{ct}(\coords) = V_{ct}(\vec{r_c},\xi)= \sqrt{4 \pi}  \sum_{Q q}
V_{Qq} (r_c,\xi)~  Y_{Qq}(\hat{r}_c) , 
\ee
where $\vec{r}_c=\vec{R}-a\vec{r}$ (see Fig. \ref{fig:1}), with
$a=m_v/(m_v + m_c)$ ($m_c$ 
and $m_v$ denote the core and valence particle masses, respectively).

In some models, such as in the rotational model assumed here, the
multipole terms $V_{Q q} (r_c,\xi)$ factorize into a radial part and a
structure part, i.e.,  
\be
\label{vct2}
V_{ct}(\coords) = \sqrt{4 \pi}  \sum_{Q q} {\cal V}^{Q}_{ct}(r_c)~
{\cal T}^{*}_{Q q}(\xi)~ Y_{Q q}(\hat{r}_c) . 
\ee
Note that $V_{ct}(\coords)$ will contain, in general, both Coulomb and
nuclear parts so, in this formalism, both interactions are  treated
simultaneously.  
%In the case of Coulomb interaction, as we assume a structureless target and a neutron valence particle, the potential 
%is limited to the sum with all the protons of the core, i.e.:
The matrix elements (\ref{cpot}) were explicitly evaluated in
Ref.~\cite{Summers06}, giving rise to the expression 
\begin{align}
\label{vjt}
U^{J_T}_{\beta:\beta'}(R) &=
\hat{L}\hat{L'}\hat{J_p}\hat{J_p^\prime}
(-1)^{J_p+J_T} \sum_{\Lambda} (-1)^{\Lambda} \hat{\Lambda}^2 
\nonumber \\ 
&\times
\threejz{\Lambda}{L}{L'} \sixj{J_p}{J_p^\prime}{\Lambda}{L'}{L}{J_T}
F^{\Lambda}_{J_p n:J_p' n'}(R) \ .
\end{align}

The form factors, $F^{\Lambda}_{J_p n:J_p' n'}(R)$ are given by
%the sum over $KQ\lambda$ multipoles and the projectile coupled channels ($a$ and $a'$), 
\be
\label{formf}
%F^{\Lambda}_{J_p n:J_p' n'}(R) = \sum_{KQ\lambda\atop \alpha, \alpha'}
F^{\Lambda}_{J_p n:J_p' n'}(R) = \sum_{KQ\lambda  \alpha, \alpha'}
{\cal R}^{KQ\lambda}_{\alpha n: \alpha' n'}(R)~ P^{KQ\lambda:\Lambda}_{\alpha:\alpha'} \ ,
\ee
with the radial integral:
\begin{align}
\label{radial}
{\cal R}^{KQ\lambda}_{\alpha n: \alpha' n'}(R)  & = \hat{K} \int
u_{n,\alpha}^{J_p*}(r) {\cal V}^{QK}_{ct}(r,R)  \nonumber \\
 & \times R^{\lambda} (a r)^{Q-\lambda} u_{n',\alpha'}^{J_p^\prime}(r) \dif r \ ,
\end{align}
where
\be
{\cal V}^{QK}_{ct}(r,R) = \frac{1}{2} \int_{-1}^{+1} \frac{{\cal V}^Q_{ct}(r_c)}{r_c^{Q}} P_K(u)
\dif u \,; \,\, u=\hat{R} \cdot\hat{r} \ .
\ee

The coefficients $P^{KQ\lambda:\Lambda}_{\alpha:\alpha'}$ are explicitly written as
\begin{align}
\label{pcoup}
P^{KQ\lambda:\Lambda}_{\alpha:\alpha'} & =
 (-1)^{j'+\ell+\ell'+s+Q} 
\hat{Q}^2 \hat{K}\hat{j}\hat{j'}\hat{\ell}\hat{\ell'} 
\nonumber \\
&\times  \threejz{K}{\lambda}{\Lambda} \sqrt{\frac{(2Q)!}{(2\lambda)![2(Q-\lambda)]!}} \reduced{I}{{\cal T}_{Q}(\xi)}{I'}
\nonumber \\ 
&\times \sum_{\Lambda'} \hat{\Lambda}'^2
\threejz{K}{Q-\lambda}{\Lambda'} \threejz{\Lambda'}{\ell}{\ell'} 
\nonumber \\
&\times \sixj{\Lambda'}{\Lambda}{Q}{\lambda}{Q-\lambda}{K}
\sixj{j}{j'}{\Lambda'}{\ell'}{\ell}{s}
\ninej{J_p}{J_p'}{\Lambda}{j}{j'}{\Lambda'}{I}{I'}{Q} \ ,
\end{align}
which, in addition to geometric coefficients, contain the structure
reduced matrix elements $\reduced{I}{{\cal T}_{Q}(\xi)}{I'}$.  
Specific models enter into these expressions through the radial form
factors ${\cal V}^{Q}_{ct}(r_c)$ and the structure reduced matrix
elements. We give explicit expressions of these magnitudes for the
model used in the calculations presented in this work.  

For the Coulomb part of the core-target interaction, we use the usual
multipole expansion  
\begin{align}
V^\mathrm{coul}_{ct}(\vec{r}_c,\xi) 
&=  \sum_{Q,q}  \frac{4 \pi}{2Q+1}~ \frac{Z_t e}{r^{Q+1}_c}~  {\cal M}(EQq)~
Y_{Q q}(\hat{r}_c)  , 
\label{vcou}
\end{align}
where ${\cal M}(EQq)$ is the multipole electric operator. Comparing with the
general expression (\ref{vct}) we have 
\be
V_{Qq}(r_c,\xi)  \equiv \frac{\sqrt{4 \pi}}{2Q+1}~ \frac{Z_t
  e}{r^{Q+1}_c}~  {\cal M}(EQq) . 
\ee

% nuclear 
For the nuclear part of the core-target interaction, we follow the
same approach as for the valence-core interaction, that is,  
we start with a central interaction, $V_{ct}^{(0)}(r_c)$, that is
deformed assuming a quadrupole deformation characterized by a  
deformation length $\delta_2$.  
%This deformed potential is expanded in multipoles and transformed
%from the intrinsic frame to the laboratory frame. This gives rise to
%a result analogous to that of Eq.~(\ref{vdef_vc}) 
%that can be casted in the form of Eq.~(\ref{vct2}) \cite{Tam65}
%-----------------------------------------------------------------------------------------
\begin{comment}
Regarding the nuclear part, we use  the particle-rotor model (PRM)
described, 
 for instance, in \cite{BM}, so we assume a core with a permanent 
 deformation. For the sake of simplicity, the discussion here 
will be restricted to a quadrupole and axially symmetric
deformation. Therefore, it can be characterized by a single parameter
$\beta_2$ and the surface radius, referred to the body-fixed frame, 
is read as $R_c(\theta',\phi')=R_{c_0} (1+\beta_2
Y_{20}(\theta',\phi'))$, with $R_{c_0}$ an average radius of the
core. 

We also assume that the core-target potential depends on the distance
between the deformed core and the target, so we can obtain it by
deforming a central potential $V_{ct}^{(0)}(r_c)$, i.e.,  
$V_{ct}^\mathrm{nuc}(r_c,\theta',\phi')=V_{ct}^{(0)}\left[r_c-\delta_2
  Y_{20}(\theta',\phi')\right]$, 
with $\delta_2=\beta_2 R_{c_0}$ being the deformation length. 
\end{comment}
%-------------------------------------------------------------------------------------------
The resulting potential is expanded in spherical harmonics and
transformed to the laboratory frame, giving rise to the result
analogous to that of Eq.~(\ref{vdef_vc}) 
%that can be casted in the form of Eq.~(\ref{vct2}) \cite{Tam65}
\begin{align}
V_{ct}^\mathrm{nuc}(r_c,\theta,\phi)&= \sqrt{4 \pi} \sum_{Q q} {\cal V}^{Q}_{ct}(r_c)~
{\cal D}^{Q}_{q 0}(\alpha',\beta',\gamma')~ Y_{Q q}(\hat{r}_c) 
\label{vcoup_lab}
\end{align}
with $\{\alpha',\beta',\gamma'\}$ the corresponding Euler angles and the radial form factors
\be
{\cal V}^{Q}_{ct}(r_c) =  \frac{\hat{Q}}{2} \int_{-1}^{1}
V_{ct}^\mathrm{nuc}\left(r_c-\delta_2 Y_{20}(\theta',0)\right) P_{Q}(u)~ \dif u , 
\ee
with $u=\cos(\theta')$.
% and ${\cal D}^{Q}_{q 0}(\alpha,\beta,\gamma)$ is a rotation matrix,
% depending on the Euler angles $\{\alpha,\beta,\gamma\}$ which define
% the transformation from the body-fixed frame to the laboratory
% frame.  
Comparing with Eq.~(\ref{vct2}) we have 
$ {\cal T}_{Qq} \equiv {\cal D}^{Q*}_{q 0}$. The reduced matrix
elements of this operator are to be calculated between rotational
states belonging to a rotational band characterized by a projection
along the symmetry axis $K$. Explicitly (see e.g.~\cite{Tho09}): 
\begin{equation}
%\langle K I \| {\cal T}_Q \| K I' \rangle =
 \langle K I \| {\cal D}^{Q*} \| K I' \rangle =
\hat{I}'  \langle I' K  Q 0 | I K \rangle ,
\label{tred_nuc}
\end{equation}
where the convention of Bohr and Mottelson~\cite{BM} for reduced matrix elements
has been assumed.

%--------------------------------------------------------------
\section{\label{sec:calc} Application to \nuc{11}{Be} reactions}
%---------------------------------------------------------------

%\subsection{Energy spectrum and wave functions in the PS basis}
%We apply the developed formalism to describe the structure of the $^{11}$Be nucleus. 

As an illustration of the formalism presented in the preceding
sections, we consider the scattering of the halo nucleus $^{11}$Be on
$^1$H, $^{64}$Zn and $^{208}$Pb targets, comparing with available  
data for these reactions. The bound and unbound states of the
$^{11}$Be  are known to contain significant admixtures  
of core-excited components \cite{For99,Win01,Cap01}, and hence core
excitation effects are expected to be important. This has been in fact  
confirmed in the case of resonant breakup of this nucleus on $^1$H
\cite{Cre11,Mor12} and $^{12}$C targets \cite{Mor12b}, using the
no-recoil XDWBA method.

As in previous works \cite{Cre11}, the  $^{11}$Be structure is
described with the particle-rotor model of Bohr and Mottelson with the
Hamiltonian of Ref.~\cite{Nun96} (model Be12-b), which consists of a 
Woods-Saxon central part, with a fixed geometry ($R=2.483$~fm,
$a=0.65$~fm) and a  parity-dependent strength ($V_{c}=-54.24$~MeV for
positive-parity states and $V_{c}=-49.67$~MeV for negative-parity ones). The
potential contains also a spin-orbit term, whose radial dependence is
given by the derivative of the central Woods-Saxon part, and strength
$V_{so}=8.5$~MeV. For the $^{10}{\rm Be}$ core, this model assumes a 
permanent quadrupole deformation $\beta_2$=0.67, (i.e.~$\delta_2=\beta_2 R$=1.664~fm). Only the ground state
($0^+$) and the first excited state ($2^{+}$, $E_x= 3.368$ MeV) are
included in the model space. For the valence-core orbital angular
momentum, we consider the values $\ell \leq 3$.  

The particle-rotor model that we assumed may seem unrealistic, but it has been proved to provide a reasonable description of $^{11}$Be \cite{Cre11,Tar04,Tar06} and other nuclei, such as several odd carbon isotopes \cite{Summers06,Tar04,Kar08}. More realistic descriptions of $^{11}$Be have indeed been proposed in the literature (see e.g.~Refs.~\cite{Kan02,For05,Lay14}) but the use of these more sophisticated structure models would make the evaluation of the coupling potentials much more involved (this can be seen for example in the microscopic cluster model recently applied to the description of the scattering of $^{7}$Li~\cite{Des13}). For the purpose of the present work, we believe that the assumed rotor model provides a simple but still reliable choice.

To generate the THO basis for all the studied cases here, we use the LST of Eq.~(\ref{lst}) with
$m=4$ and $b=1.6$~fm.  The number of oscillator functions N, and the
$\gamma$ parameter are determined specifically for each reaction, and
will be specified below.

%---------------------------------------------------------------
%\subsection{Application to ${\rm ^{11}B\MakeLowercase{e}}+
%\MakeLowercase{p}$ resonant breakup \label{sec:calc}}  
\subsection{Application to $^{11}$Be+ $p$ resonant breakup \label{sec:be11p}} 
%--------------------------------------------------------------------
We first apply the XCDCC + THO method to  the  breakup
of $^{11}$Be on a proton target at 63.7 MeV/nucleon and compare with
the data of \cite{Shr04}. The measured data consist in angular
distributions for two intervals of the  neutron$-^{10}$Be core energy:
(i) $E_{\rm rel}$=0--2.5 MeV and (ii) $E_{\rm rel}$=2.5--5.0 MeV. The
first interval contains a narrow $5/2^+$ resonance at
$E_\mathrm{rel}$=1.28 MeV \cite{Kel12}. This resonance has a dominant
$^{10}\mathrm{Be}(0^+)\otimes \nu 1d_{5/2}$ parentage and a small
 $^{10}\mathrm{Be}(2^+)\otimes \nu 2s_{1/2}$ component. The cross
section for the second interval contains presumably contributions
coming from several resonances, namely, $E_x$=2.64~MeV ($3/2^-$),
3.40~MeV ($3/2^-$, $3/2^+$), 3.89~MeV ($5/2^-$), and 3.95~MeV
($3/2^-$) \cite{Kel12}. Previous calculations \cite{Cre11,Mor12},
based on the no-recoil XDWBA method, showed indeed that the main contribution
to the lower energy angular distribution arises from the
single-particle excitation mechanism populating the $5/2^+_1$
resonance, whereas for the higher energy angular distribution the main
contribution comes from the excitation of the $3/2^+_1$ resonance due
to the collective excitation  of the $^{10}$Be core. 

We repeat the calculations of Refs.~\cite{Cre11,Mor12} using the more
sophisticated XCDCC formalism for the reaction dynamics, and the THO
PS basis for the $^{11}$Be states. The LST was generated with the
parameter $\gamma=1.6$~fm$^{1/2}$. The number of
oscillator functions was N=14. Continuum states with $J_p=1/2^\pm$,
$3/2^\pm$ and $5/2^+$ were found to be enough for convergence of the
calculated observables. The proton-neutron interaction was represented by a simple Gaussian interaction 
derived in Ref.~\cite{Mor12}, while for the core-target potential
we used the CH89 optical model parametrization \cite{Var91}, but
 modifying the real and imaginary depths in order to reproduce the experimental available
 elastic and inelastic data of $^{10}$Be+p at 59.2~MeV/nucleon \cite{Iwa00}. In order to reproduce the magnitude of the inelastic data, this potential required a deformation length of $\delta_2=1.9$~fm, which is somewhat larger than the deformation used in our adopted rotor model for $^{11}$Be, but consistent with  the values extracted in the DWBA analysis done in \cite{Iwa00} for the same data. For the nuclear part, both monopole $Q=0$
and quadrupole $Q=2$ terms were included. For the Coulomb part, following the expression given in (\ref{vcou}), both terms are also considered even though the $Q=0$ term gives the main contribution for such a light system. The value of $\langle 0^+ \| {\cal M}(E2) \| 2^+ \rangle$  was derived from the experimental value of
$B(E2; 0^+ \rightarrow  2^+)=53(6)$~e$^2$fm$^4$ \cite{Ram87}. The reorientation term $2^+
\leftrightarrow 2^+$  was also included, and the value of the reduced
matrix element $\langle 2^+ \| {\cal M}(E2) \| 2^+ \rangle$ was derived from
the computed value of $\langle 0^+ \| {\cal M}(E2) \| 2^+ \rangle$, assuming
that the $0^+$ and $2^+$ states of $^{10}$Be are members of the same
rotational band with $K^\pi=0^+$. %only
%the $Q=0$ term was included since higher multipoles were found to be
%negligible for such a light system.

\begin{figure}
{\par\centering \resizebox*{0.45\textwidth}{!}{\includegraphics{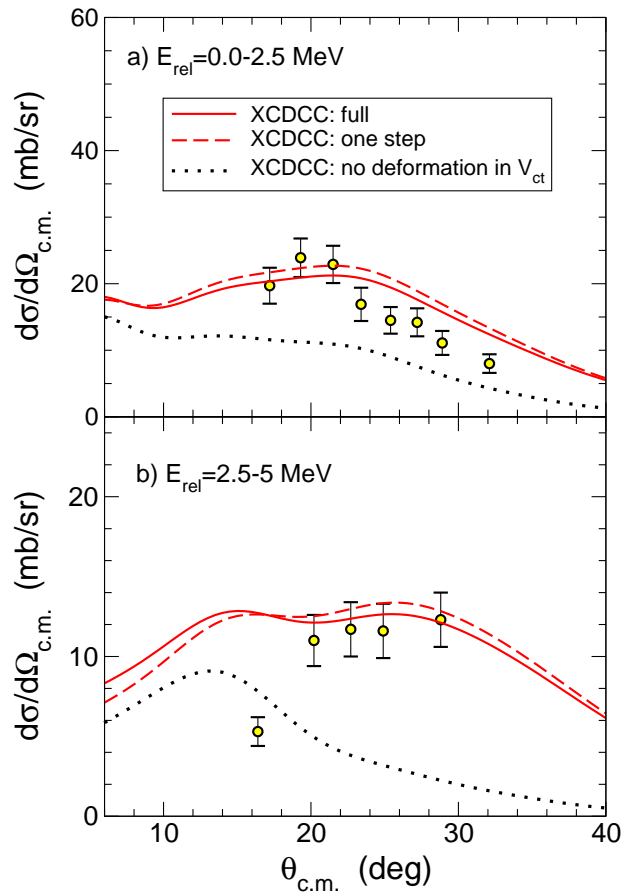}}\par}
 \caption{\label{be11p} (Color online) Differential breakup cross
   sections, with respect to the outgoing $^{11}$Be$^{*}$ c.m.\  
scattering angle, for the breakup of $^{11}$Be on protons at 63.7
MeV/nucleon. Upper and bottom panels correspond to the neutron-core
relative energy intervals $E_\mathrm{rel}$=0--2.5~MeV and
$E_\mathrm{rel}$=2.5--5~MeV, respectively.}  
\end{figure}

The results of these calculations are shown in Fig.~\ref{be11p}. The
solid line corresponds to our full coupled-channels calculation,
including couplings to all orders. Considering the experimental error bars, the agreement with
the data is fairly reasonable for both energy intervals, except for
the first data point in the higher energy interval. These results are qualitatively similar 
to those found in Refs.~\cite{Cre11,Mor12} using the no-recoil XDWBA approach.

To illustrate the importance of higher-order effects, we include also
the first-order calculation assuming a one-step breakup mechanism
(dashed lines). Differences between the latter and the full
calculations are small but not negligible, indicating that, even at
these relatively high incident energies, higher order effects are
significant and hence an accurate description of these reactions
require going beyond the simpler DWBA approximation.  
We also include the results obtained omitting the $Q=2$ term in the
nuclear part of the core-target interaction, but keeping the
deformation in the neutron-core interaction (dotted line). This result
differs significantly from the full calculation, indicating very
clearly the sizable effect of the dynamic core excitation mechanism
during the collision. The difference is particularly noticeable in the higher energy interval, due to fact that this interval 
is dominated by the $3/2^+$ resonance, which is mostly populated by a core excitation mechanism \cite{Cre11}.

It is worth noting that our results differ qualitatively from those performed in Ref.~\cite{Sum07} for the same reaction, using also the XCDCC formalism, but with a binning discretization scheme. In benchmarking the results of the present work with those from \cite{Sum07} several mistakes were found in the equations of that reference, as well as in their numerical implementation. These mistakes result in a significant underestimation of core excitation effects \cite{Sum14,Sum14b}. In addition, some differences are expected due to the different choice of the p-n interaction.

%--------11Be+p dsdw ---------------
%------------------------------------

\begin{figure}
{\par\centering \resizebox*{0.45\textwidth}{!}{\includegraphics{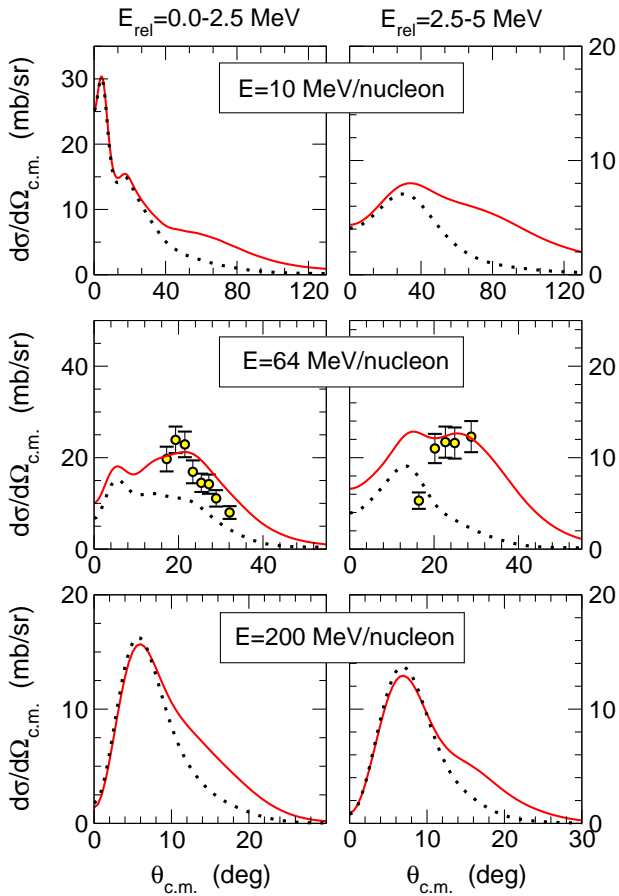}}\par}
 \caption{\label{be11p_edep} (Color online) Energy dependence of the core excitation effects in the breakup of 
$^{11}$Be on protons. The top, middle and bottom rows correspond to the bombarding energies 10, 64 and 200 MeV/nucleon, respectively. 
The left and right panels are for the neutron-core energy intervals $E_\mathrm{rel}$=0--2.5~MeV and
$E_\mathrm{rel}$=2.5--5~MeV, respectively. In each panel, the solid line is the full XCDCC calculation, whereas the dotted line is the XCDCC calculation without deformation in the $p$+$^{10}$Be potential.}  
\end{figure}

We have studied also the dependence of the core excitation effect with the incident energy. For this, 
%Intuitively, one would expect that the core excitation mechanism becomes more important as the incident energy of the projectile increases. To illustrate this energy dependence,
 we have performed additional XCDCC calculations for the same reaction at 10 MeV/nucleon and  200 MeV/nucleon. At both energies, we use  the 
CH89 parametrization for the $^{10}$Be+p  interaction. Because of the lack of experimental data at those energies, we keep the deformation length obtained from the fit of the inelastic data performed at 59.2 MeV ($\delta_2=1.9$~fm), and adjust the potential depths in order to reproduce the elastic and inelastic scattering of $^{10}$Be+$p$  obtained with a microscopic folding potential, generated with the JLM interaction and transition densities from antisymmetrized molecular dynamics (AMD) calculations (see Ref.~\cite{Tak08} for a similar approach).
%    For these test calculations, we kept the same potentials, and varied only the incident energy. 
The results are shown in Fig.~\ref{be11p_edep}. 
The calculations shown in the middle panels are just the same as those shown in Fig.~\ref{be11p}, but are included here to facilitate the comparison using a wider angular range. One can see that the core excitation energy is important at the three incident energies, particularly in the region containing the $3/2^+$ resonance. 

% One can see that, at 10 MeV/nucleon, the contribution of core excitation to the breakup mechanism, although still present, is smaller than at 64 MeV/nucleon. On the contrary, at 250 MeV/nucleon, the core excitation contribution is remarkably large, and dominates the breakup cross section. 

%--------11Be+p dsdw ---------------
%------------------------------------

%-------------------------------------------------------------------------------------
\subsection{Application to $^{11}$Be+ $^{64}$Zn elastic and breakup \label{sec:be11zn}} 
%-------------------------------------------------------------------------------------
As a second example, we consider the $^{11}$Be+ $^{64}$Zn reaction at
28.7 MeV. Quasi-elastic (elastic + inelastic) and inclusive breakup
data from this reaction have been reported in Ref.~\cite{Pie10} and
have been analyzed within the standard CDCC framework in several works
\cite{Kee10,Pie12,Dru12}. 

We compare these data with XCDCC calculations. For the neutron-target
interaction, we used the same optical potential used in
Ref.~\cite{Pie12}. For the $^{10}$Be+ $^{64}$Zn interaction, we
started from the optical potential derived in Ref.~\cite{Pie10}  from
a fit of the elastic scattering data for this system. To account for
the core excitation mechanism, this potential is deformed with the
same deformation length used in the structure model, i.e.,
$\delta_2=1.664$~fm. Coulomb multipoles (monopole and quadrupole) were also included, as in the previous case, according
to the expansion (\ref{vcou}). %The value of $\langle 0^+ \| {\cal M}(E2) \|
%2^+ \rangle$  was derived from the experimental value of
%$B(E2; 0^+ \rightarrow  2^+)=53(6)$~e$^2$fm$^4$ \cite{Ram87}. The reorientation term $2^+
%\leftrightarrow 2^+$  was also included, and the value of the reduced
%matrix element $\langle 2^+ \| {\cal M}(E2) \| 2^+ \rangle$ was derived from
%the computed value of $\langle 0^+ \| {\cal M}(E2) \| 2^+ \rangle$, assuming
%that the $0^+$ and $2^+$ states of $^{10}$Be are members of the same
%rotational band with $K^\pi=0^+$.  
To recover the description of the $^{10}$Be+ $^{64}$Zn elastic data,
once these additional couplings are included, the optical potential
depths were readjusted, giving rise to the modified values
$V_0=-84.5$~MeV and $W_v=-34.1$~MeV, for the real and imaginary parts,
respectively.

For the $^{11}$Be projectile, continuum states up to $J_p=7/2$ (both
parities) were included. These states were obtained by diagonalizing
the $^{11}$Be Hamiltonian in a THO basis with $N=10$ radial functions,
$\ell \leq 3$ and $I=0,2$. For the LST, the parameter
$\gamma=1.8$~fm$^{1/2}$ was used, although additional tests were done
with other choices to verify the independence and stability of the
results with respect to parameters $b$ and $\gamma$. After diagonalization, only
eigenstates below 13 MeV were retained for the coupled-channels
calculations. We verified that including eigenstates up to 14 MeV  had a very small effect on the studied observables.  

\begin{figure}
{\par\centering \resizebox*{0.45\textwidth}{!}{\includegraphics[angle=0]{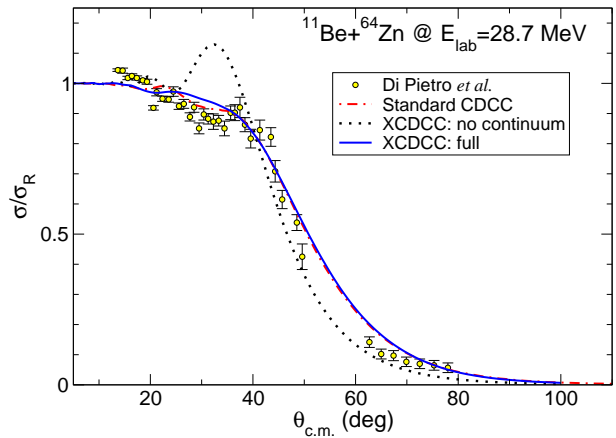}}\par}
 \caption{\label{be11zn_qel} (Color online) Quasi-elastic
   differential cross section, relative to Rutherford, for the
   scattering of $^{11}$Be on $^{64}$Zn at $E_\mathrm{lab}$=28.7~MeV. The (red) dot-dashed
   line is the standard CDCC calculation, without core excitation,
   from Ref.~\cite{Pie12}. The blue solid line is the XCDCC
   calculation. The dotted line is the XCDCC calculation neglecting
   the coupling to the breakup channels. Experimental data are from
   Ref.~\cite{Pie10}.}   
\end{figure}

The calculated differential quasi-elastic cross section is compared
with the data in Fig.~\ref{be11zn_qel}. The dotted line is the XCDCC
calculation neglecting the coupling to the breakup channels, that is,
including only the $^{11}$Be ground-state and first excited state. As
expected, this calculation largely fails to describe the data. The
solid line is the full XCDCC calculation. This calculation describes
well the data in the full angular range. We have also included the result obtained 
with the standard CDCC calculation from Ref.~\cite{Pie12}. Except for some small differences around  $\theta_\mathrm{c.m.}\approx 30^\circ$ turn out 
to be very similar. 

%However, this similitude in the quasi-elastic cross section does not imply an agreeme
Although the data of Ref.~\cite{Pie10} did not provide the separate
contribution of the inelastic cross section for the $1/2^-$ bound state
at $E_x=320$~keV, it is worth comparing the values computed with the two methods. In the CDCC calculations of
Ref.~\cite{Pie12}, the total inelastic cross section for the
population of this state was about 750~mb, whereas in the XCDCC
calculation this value is reduced to $\sim$566~mb. The difference can
be understood comparing the values of the electric transition probability $B(E1;\mathrm{g.s.} \rightarrow
1/2^-)$ for these two models. The single-particle model used in
Ref.~\cite{Pie12} yields  $B(E1;\mathrm{g.s.} \rightarrow
1/2^-)=0.260$~e$^2$fm$^2$, whereas for the PRM model used here it is
$B(E1;\mathrm{g.s.} \rightarrow 1/2^-)=0.140$~e$^2$fm$^2$. This value
is in better agreement with the experimental one, $B(E1;\mathrm{g.s.}
\rightarrow 1/2^-)=0.116$~e$^2$fm$^2$ \cite{Mil83}, so we expect that the inelastic cross section calculated with XCDCC be 
more realistic than that obtained with the standard CDCC method.

%--------11Be+64Zn ---------------
%------------------------------------

\begin{figure}
{\par\centering \resizebox*{0.45\textwidth}{!}{\includegraphics[angle=0]{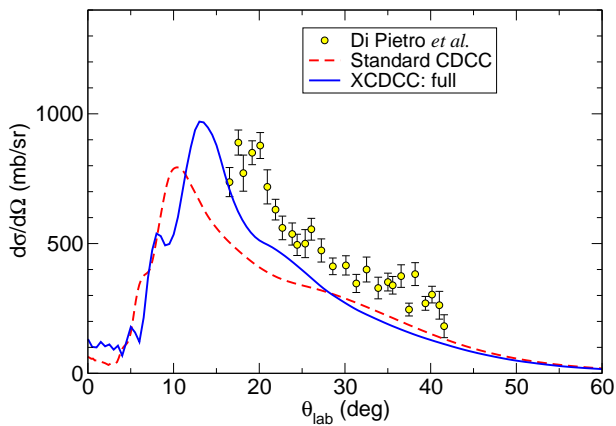}}\par}
 \caption{\label{be11zn_bu} (Color online) Differential cross
   section, as a function of the laboratory angle, for the $^{10}$Be
   fragments resulting from the breakup of  $^{11}$Be on $^{64}$Zn at
   $E_\mathrm{lab}$=28.7~MeV. The (red) dashed line is the standard
   CDCC calculation, without core excitation, from
   Ref.~\cite{Pie12}. The blue solid line is the XCDCC
   calculation. Experimental data are from Ref.~\cite{Pie10}.}  
\end{figure}

The same experiment also provided the inclusive $^{10}$Be angular
distribution.  In
Fig.~\ref{be11zn_bu} we compare the data from Ref.~\cite{Pie10} with
the present XCDCC calculations and the standard CDCC
calculations from Ref.~\cite{Pie12}. It is worth noting that the data
are referred to the $^{10}$Be laboratory angle. The calculation of
this observable within the XCDCC framework would require an appropriate
kimematical transformation, similar to that developed in
Ref.~\cite{Tos01} for the standard CDCC method, but this formalism is
not yet available for XCDCC. Consequently, we perform an approximate
transformation, approximating the $^{10}$Be scattering angle by the
$^{11}$Be$^*$ scattering angle.  

%--------11Be+64Zn ---------------
%------------------------------------

The XCDCC calculation is found to be larger than the CDCC result. This
increase improves the agreement with the data of Ref.~\cite{Pie10} although
some underestimation is still observed. This remaining discrepancy could
be due to the limitations of the $^{11}$Be model used in the XCDCC
calculations, but also to the contribution of non-elastic breakup events in
the data.  It is worth noting that the CDCC and XCDCC methods provide only the so-called elastic breakup component, that is,  the projectile
dissociation in which both the neutron and core survive and the target is left in the ground state. However, since the neutrons were not detected in the experiment of Ref.~\cite{Pie10}, the data might contain also contributions from other processes involving the absorption of the neutron by the target and/or the target excitation.  
%,  since the neutrons were not detected in this experiment
%the $^{10}$Be events might arise from projectile breakup, but also from a
%one-neutron transfer mechanism. The CDCC method provides only the
%so-called elastic breakup cross section, that is, the projectile dissociation leaving the target in the ground state. 

%---------- 11Be+64Zn convergence with Jp_max 
%\begin{figure}
%{\par\centering \resizebox*{0.45\textwidth}{!}{\includegraphics{be11zn_jmax.eps}}\par}
%\caption{\label{be11zn_jmax} (Colour online) Convergence of the
%quasi-elastic (upper figure) and breakup (bottom figure) cross
%sections with respect to the projectile angular momentum for the
%reaction $^{11}$Be on $^{64}$Zn at $E_\mathrm{lab}$=28.7~MeV. {\bf
%AMM: ESTA FIGURA NO ESTA COMENTADA EN EL TEXTO, PORQUE NO SE SI
%MERECE LA PENA DEJARLA O NO}  }   
%\end{figure}

\begin{figure}
{\par\centering \resizebox*{0.45\textwidth}{!}{\includegraphics{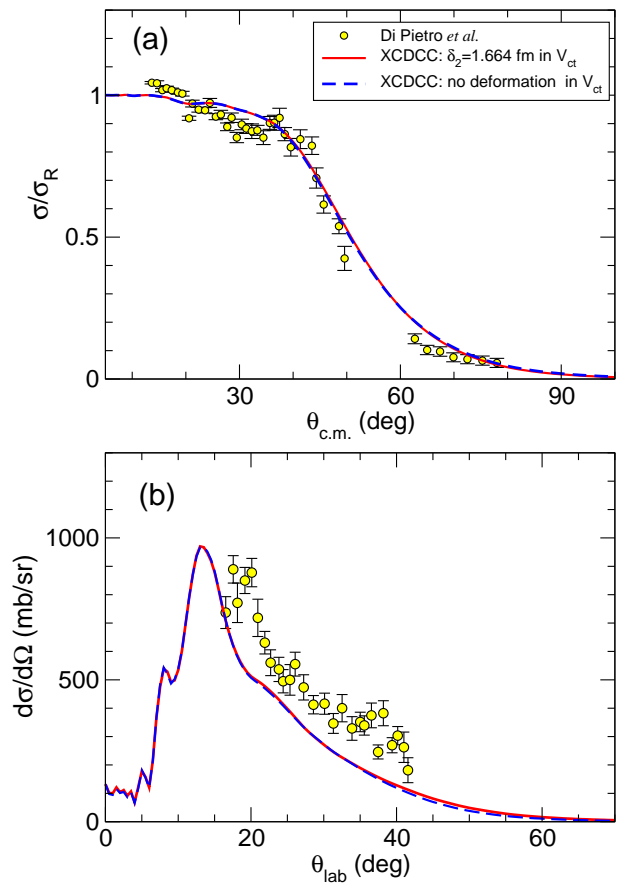}}\par}
 \caption{\label{be11zn_vcdef}(Color online) Quasi-elastic
   differential cross section (top) and breakup differential cross
   section (bottom) for $^{11}$Be on $^{64}$Zn at
   $E_\mathrm{lab}$=28.7~MeV, compared with XCDCC calculations, for
   different choices of the $^{10}$Be+$^{64}$Zn potential.  See text for further details.}  
\end{figure}

%--- vct
In the case of $^{11}$Be+$p$ breakup at intermediate energies, we found
that the deformed part of the core-target interaction gives rise to an
increase of the breakup cross sections. Now we study the effect
of these  terms  in the
$^{11}$Be+$^{64}$Zn case. For this purpose, we compare in
Fig.~\ref{be11zn_vcdef} the full XCDCC calculation, described above,
with another XCDCC calculation, in which the $^{10}$Be+$^{64}$Zn
potential is described with the central optical potential of
Refs.~\cite{Pie10,Pie12}.  Both calculations give
almost identical results for the quasi-elastic and breakup data. This
result indicates that, at these low incident energies (a few MeV per
nucleon) and for medium-mass targets, the dynamic core excitation effect due to the core-target potential is
well represented by an optical potential describing the corresponding
elastic data. Consequently, at these energies, the main effect of core
excitation comes from the admixtures of core-excited components in the
projectile wave functions.  

%--------11Be+64Zn for Vct
%------------------------------------

%-------------------------------------------------------------------------------------
\subsection{Application to $^{11}$Be+ $^{208}$Pb  breakup \label{sec:be11pb}} 
%-------------------------------------------------------------------------------------
As a final example, we consider the reaction of $^{11}$Be on a
$^{208}$Pb target. This reaction has been measured by several groups
\cite{Nak94,Pal03,Fuk04} at intermediate energies (several tens of MeV
per nucleon) with the aim of obtaining information on the dipole
Coulomb response of $^{11}$Be as well as on the amount of $s$-wave
component in the ground state. We have performed XCDCC calculations at  
69 MeV/u, which corresponds to the energy of the experiment performed
at RIKEN by Fukuda \etal \cite{Fuk04}.  

Continuum states with $J_p$=$1/2^\pm$, $3/2^\pm$ and $5/2^+$ were
considered (test calculations revealed that the effect of the $5/2^-$
states is negligible for the studied angles). These  states were
generated with a THO basis with $N=15$ states and
$\gamma=1.8$~fm$^{1/2}$ ($b=1.6$~fm as in the preceding cases). After
diagonalization, all eigenstates below 
8~MeV were retained for the coupled-channels calculation.  

For the neutron-target interaction we used the parametrization of
Koning and Delaroche \cite{KD03}. The central part of the core-target
potential was taken from Ref.~\cite{Cap03} (first line of Table
III). As in the previous case, this potential is deformed with a
deformation length of $\delta_2=1.664$~fm. At these relatively high energies the
breakup process is  essentially a one-step mechanism connecting the
ground state directly with the breakup channels. Moreover, at the very forward angles
measured in the experiment of Ref.~\cite{Fuk04} one expects that the breakup is largely dominated by the dipole
Coulomb couplings.   Consequently, at
these angles the most strongly coupled breakup states will be the $1/2^-$
and $3/2^-$.  These states cannot be populated by the dynamic core
excitation mechanism in first order since the quadrupole nature of
these excitation connects the ground state with positive parity
continuum states. As a result, for this reaction (and in general for
other reactions induced by weakly-bound nuclei on heavy targets) the
main core-excitation effect is due to the presence of core-excitation
admixtures in the projectile states.   

%----------------------- 11Be+208Pb-----------------------------------------------------
\begin{figure}
{\par\centering \resizebox*{0.45\textwidth}{!}{\includegraphics[angle=-90]{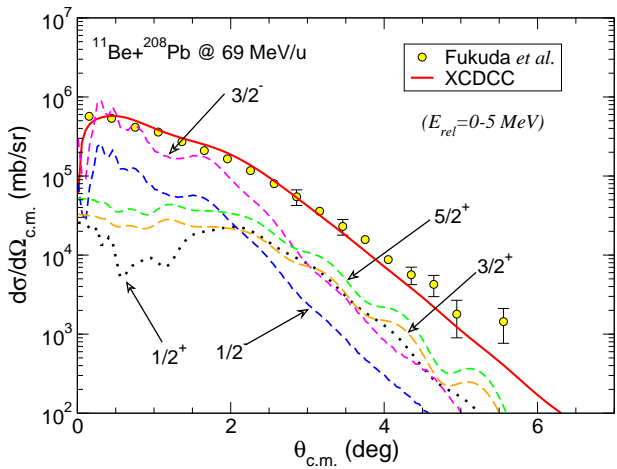}}\par}
 \caption{\label{be11pb_dsdw} (Color online)  Breakup differential cross section  for  
$^{11}$Be on $^{208}$Pb at $E_\mathrm{lab}$=69~MeV/nucleon, integrated in the
   n-$^{10}$Be relative energy up to 5 MeV. The data are from
   Ref.~\cite{Fuk04}. The lines are XCDCC calculations described in
   the text. The full calculation (solid line) has been convoluted
   with the experimental angular resolution for a meaningful
   comparison with the data.}   
\end{figure}
%-----------------------------------------------------------------------------------------

The XCDCC calculations are compared with the data of Ref.~\cite{Fuk04}
in Fig.~\ref{be11pb_dsdw}. The solid line is the full
calculation. Since the experimental distribution was integrated for
relative $n$-$^{10}$Be energies below 5 MeV, the theoretical result was
obtained adding the angular distributions for positive-energy
pseudo-states lying below this energy. The final
distribution was  convoluted with the experimental energy resolution
quoted in \cite{Fuk04}. This calculation is found to be in very good
agreement with the data (solid line in Fig.~\ref{be11pb_dsdw}). It is worth noting that no scaling factor is
introduced in the calculation. The analysis of this kind of
experiments is usually done assuming single-particle states for the
initial and final states. The final result is then renormalized by a
scaling factor which, in the present case, can be interpreted as the spectroscopic factor for
the  $^{10}{\rm Be} \otimes s_{1/2}$ configuration in the ground-state
wavefunction. In our calculations, this spectroscopic factor is
already included in the description of the ground-state wavefunction.  

To illustrate the dominance of the dipole excitation mechanism, we
have plotted also in Fig.~\ref{be11pb_dsdw} the separate contribution of the $1/2^\pm$, $3/2^\pm$
and $5/2^+$ states.   It can be seen that, at sufficiently small angles, the breakup
is largely dominated by the coupling to the dipole states
and, in particular, to the $3/2^-$ states.

\begin{figure}
{\par\centering \resizebox*{0.45\textwidth}{!}{\includegraphics[angle=0]{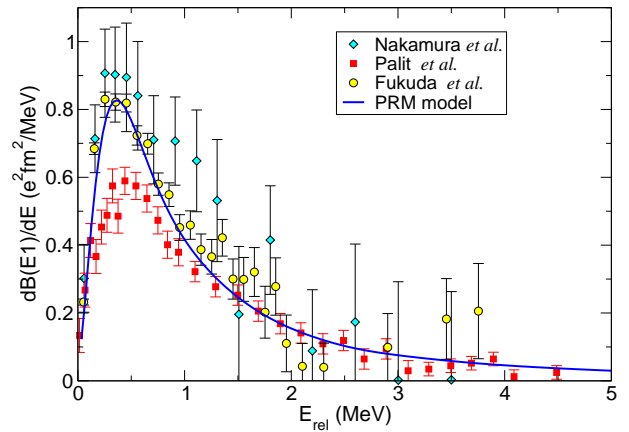}}\par}
 \caption{\label{dbde_be11} (Color online)  Dipole strength
   distribution for $^{11}$Be deduced from Coulomb breakup
   experiments: diamonds \cite{Nak94}, squares \cite{Pal03}, circles
   \cite{Fuk04} and from the PRM model of Ref.~\cite{Nun96}. The
   latter has been convoluted with the energy resolution corresponding
   to the experiment of Ref.~\cite{Fuk04}. }   
\end{figure}

This dominance of the dipole Coulomb couplings supports the procedure followed in Ref.~\cite{Fuk04} to extract the $B(E1$) response of the $^{11}$Be nucleus from the analysis of these exclusive breakup data. In Ref.~\cite{Fuk04} this was done comparing the breakup data with first-order semiclassical calculations. The extracted $B(E1)$ distribution, quoted from Ref.~\cite{Nak12}, is compared in Fig.~\ref{dbde_be11} with the theoretical  $B(E1)$ distribution obtained with the PRM model adopted in our XCDCC calculations (solid line).  We include also the {\it experimental} distributions from Refs.~\cite{Nak94,Pal03} deduced from similar Coulomb dissociation experiments. It is seen that the theoretical $B(E1)$ distribution agrees very well with the {\it experimental} $B(E1)$ distribution from  Ref.~\cite{Fuk04}, and this explains also the good agreement in the corresponding breakup cross sections. 
 
\begin{comment}
The good agreement observed in Fig.~\ref{be11pb_dsdw} between the XCDCC calculations and the data
indicates that the dipole strength probability  predicted by the
structure  model used to describe the $^{11}$Be states  is consistent
with that  extracted from the data of Ref.~\cite{Fuk04}. This is
confirmed in Fig.~\ref{dbde_be11}, where we compare the theoretical
$dB(E1)/dE$ distribution with that inferred from the analysis of
Ref.~\cite{Fuk04}, as reported in Ref.~\cite{Nak12}. We include also
the experimental distributions from Refs.~\cite{Nak94,Pal03} deduced
from similar Coulomb dissociation experiments. It is seen that the
calculated dipole response agrees well with that extracted from the
data of Ref.~\cite{Fuk04} and this is consistent with the agreement in
the differential breakup cross sections.  
\end{comment}

%----------------------- --------- dB(E1) -------------------------------------------
%--------------

%--------------------------------------------------------------------
\section{\label{sec:summary} Summary and conclusions}
%--------------------------------------------------------------------

To summarize, we have studied the scattering of a two-body halo
nucleus (core plus a valence particle) on an
inert target within an extended version  
of the Continuum-Discretized Coupled-Channels (XCDCC) formalism. The
method takes into account the effect of core excitation in  
the structure of the projectile, by allowing the inclusion of
core-excited components in the projectile states, and also in the
dynamics of the reaction, by allowing core excitation and deexcitation
during the collision.  

The projectile states are described in the weak-coupling limit. Thus,
the states of the composite system are expanded as a superposition of
products of single-particle configurations and core states. The
energies and wavefunctions of the projectile are calculated using the
pseudo-state (PS) method, that is, diagonalizing the model Hamiltonian
in a basis of square-integrable functions.  For the relative motion
between the valence particle and the core, we use the analytical
Transformed Harmonic Oscillator (THO) basis used in  previous works
\cite{Lay12}. 
% with a recently proposed extension of the analytical Transformed Harmonic Oscillator (THO) basis, 
%Following our previous choice for non-deformed systems, we propose to
%use as PS basis the Transformed Harmonic Oscillator (THO)
%basis. The basis functions are obtained by applying an analytic local
%scale transformation \cite{Amos,Mor09} to the conventional HO basis.
%The transformation is such that it converts the Gaussian asymptotic
%behavior of the HO function into an exponential.   

The method has been applied to the scattering of $^{11}$Be on several
targets. The $^{11}$Be nucleus is described in a simple  
particle-rotor model, in which the $^{10}$Be core is assumed to  have
a permanent axial deformation with $\beta_2=0.67$ 
\cite{Nun96}. The core-target interaction is obtained by deforming a
central phenomenological potential.  

To study the dependence of core excitation on the target mass, we have
performed calculations for three different targets: $^1$H, $^{64}$Zn
and $^{208}$Pb at incident energies for which experimental data exist.  

In the $^{11}$Be+$p$ reaction, the calculations reproduce well the
breakup data from Shrivastava \etal \cite{Shr04} corresponding to an
incident energy of  64 MeV/nucleon. The XCDCC results  are
qualitatively similar to those found in previous studies, using a
no-recoil XDWBA approximation \cite{Cre11,Mor12}. In particular, we
confirm the importance of the dynamic core excitation mechanism, due
to the non-central part of the core-target interaction, for the
excitation of the low-lying $5/2^+$ and $3/2^+$ resonances. Moreover,
higher order couplings are found to be non negligible and, therefore,
should be taken into account for an accurate description of similar
reactions. In particular, inclusion of breakup beyond first order is
found to improve the agreement in the absolute cross section at
excitation energies around the $5/2^+_1$ resonance.  

The $^{11}$Be+$^{64}$Zn reaction has been studied at 28.7 MeV, for
which quasi-elastic and inclusive breakup data are available   
\cite{Pie10}. The experimental quasi-elastic cross sections are well
reproduced at all angles, except for some slight overestimation 
at  $\theta_\mathrm{c.m} \approx 30^\circ$. The XCDCC result turns out to be 
very close to the standard CDCC calculation from Ref.~\cite{Pie10}. On the other 
hand, the inclusive breakup cross sections are larger than
those found in the standard CDCC calculations, being in better
agreement with the data from Ref.~\cite{Pie10}. For this medium-mass
target, the dynamic core excitation mechanism is found to be small and
the full calculations can be simulated using a central core-target
potential fitted to the $^{10}$Be+$^{64}$Zn elastic data.   

Finally, we have presented calculations for the $^{11}$Be+$^{208}$Pb
reaction at 69 MeV/u. The calculated breakup angular distribution is
found to reproduce very well the data from Ref.~\cite{Fuk04}. For this
heavy target, and at very small angles, the breakup is dominated by
the dipole Coulomb couplings connecting the ground state with the
dipole ($1/2^-$ and $3/2^-$) continuum states. In our model, these
states cannot be populated by a direct core excitation mechanism, and
hence core excitation enters only through the admixture of different
core and valence configurations in the projectile wavefunctions. These
admixtures are nevertheless very important to account for the correct
normalization of the data.  

Summarizing, the effect of core-excitation in the structure is found
to be important for all targets. However, the {\it dynamic} core excitation
mechanism is important for light targets (for which the dipole
excitations are small compared to the  quadrupole collective
excitations of the core) at all incident energies explored here. Although 
all the calculations presented in
this work have been performed for the $^{11}$Be nucleus, we believe
that the results are extrapolable to other weakly-bound nuclei and,
consequently, the effects discussed here should be taken in
consideration for an accurate description and interpretation of the
data.

%--------------------------------------------------------------------------
\begin{acknowledgments}
%--------------------------------------------------------------------------
 We are grateful to I.~J.~Thompson, N.~C.~Summers and F.~M.~Nunes for
 very useful discussions and feedbacks regarding the XCDCC formalism,
 and for letting us using an unpublished version of {\sc fresco} for
 comparison. This work has been partially supported by the Spanish
 Ministerio de  Econom\'ia y Competitividad and FEDER funds under projects 
 FIS2011-28738-c02-01,  FPA2009-07653,  FPA2009-08848 and  by the Spanish Consolider-Ingenio 2010 Programme CPAN
(CSD2007-00042)  and by Junta de  Andaluc\'ia (FQM160,
 P11-FQM-7632). One of us (R.D.) acknowledges support by the
 Funda\c{c}\~ao para a Ci\^encia e a Tecnologia (FCT) grant
 SFRH/BPD/78606/2011.  
\end{acknowledgments}

\bibliography{xcdcc}

\end{document}